\documentstyle[emulateapj,psfig]{article}
 
\lefthead{Agol, Jones, \& Blaes}
\righthead{Mid-IR Imaging of QSO 2237+0305}
\begin{document}

\title{Keck Mid-Infrared Imaging of the QSO 2237+0305}

\author{Eric Agol}
\affil{Physics and Astronomy Department, Johns Hopkins University,
    Baltimore, MD 21218}
\author{Barbara Jones}
\affil{Center for Astrophysics and Space Sciences, University
of California at San Diego, La Jolla, CA 92093} 
\author{Omer Blaes}
\affil{Physics Department, University of California,
Santa Barbara,  CA 93106} 
\begin{abstract}
Using the Long Wavelength Spectrometer on Keck, we have imaged the gravitationally 
lensed radio quiet quasi-stellar object (QSO) 2237+0305 at 8.9 and 11.7 $\micron$ 
for the first time.  The mid-infrared flux ratios are inconsistent with the optical 
flux ratios, but agree with the radio flux ratios and with some published gravitational 
lens models.  These flux ratios indicate that the infrared emission is not affected
by microlensing, which rules out the synchrotron emission model.  The infrared
emission is likely produced by hot dust extended on a length scale of more 
than 0.03 parsecs.  The spectral energy distribution further implies a narrow range 
of dust temperatures, suggesting that the dust may be located in a shell extending 
between $\sim 1$ and 3~pc from the nucleus, and intercepting about half of the QSO 
luminosity.
\end{abstract}

\keywords{gravitational lensing --- infrared: galaxies --- quasars: QSO 2237+0305}

\section{Introduction}

QSO 2237+0305 (z=1.695, Huchra et al. 1985), referred to hereafter as 
``the Einstein Cross,'' is the first gravitationally lensed QSO with confirmed 
variations due to lensing by the stars in the lens galaxy (z=0.0395), 
referred to as ``microlensing" (Irwin et al. 1989).
Optical variations of $\simeq 0.5$ magnitudes due to microlensing are
observed at least once per year (Racine 1992), which implies the optical
emission region is smaller than an Einstein radius, $R_{\rm E}=
1.1\times10^{17}M^{1/2}h^{-1/2}{\rm cm}$,
where $M$ is the mass of the lensing star in solar masses and $h=H_0/(100{\rm
km\,s^{-1}\,Mpc^{-1}})$.  The peak magnification and time scale of a high 
amplification event (caused by a caustic crossing the QSO) is strongly 
dependent on the size of the QSO.  An event in 1988 has been used to
limit the optical source size to about $10^{15}$ cm (Wambsganss, 
Schneider, \& Paczy\'nski 1990,
Rauch \& Blandford 1991, Wyithe et al. 2000).
 
Typically QSOs have a change in spectral slope near 1$\micron$ in the
rest frame, which may imply that different physical components contribute
longward and shortward of this wavelength (Neugebauer et al. 1987, 
Sanders et al. 1989, Barvainis 1993, Begelman 1994).  One way 
to distinguish these components is reverberation mapping:  observations of 
a few Seyfert galaxies indicate that the infrared emission region is more
extended than the optical (Barvainis 1992, Nelson 1996).  However,
there exist no such direct constraints on the size of the infrared emission 
in QSOs; recent observations indicate this region may be rather
compact, although the data are ambiguous (Neugebauer \& Matthews 1999).
Another technique for constraining the size of the emission region in QSOs
is gravitational microlensing.  Sources larger than an Einstein radius have 
smaller fluctuations than sources smaller than an Einstein radius (Kayser, 
Refsdal, \& Stabell 1986).  Despite its potential for probing the QSO
central engine, microlensing has not yet been looked for in the mid-infrared 
($\lambda > 3 \micron$).  Currently the most popular model of the infrared
(IR) spectrum in QSOs
is thermal emission from hot dust.  Evidence includes the lack of IR variability 
in most QSOs and the ubiquitous spectral dip around one micron which can
be ascribed to 
dust sublimation at a temperature of $T_s\simeq1800$K.  However, some radio quiet 
QSOs show phased IR/optical variability (Neugebauer \& Matthews 1999), and dust 
absorption features are not generally seen in the mid-IR in bright AGN (Roche et 
al. 1991).  Alternatively, the IR spectrum could be dominated by nonthermal 
synchrotron emission from the region near the black hole.

Microlensing variability in the mid-IR can place a limit on the size of the
source, distinguishing between dust and synchrotron models.  The {\it minimum} 
radius at which dust can exist near the central engine is 
\begin{equation}
R_{\rm dust} = 3\times10^{18} 
\left({L \over 2\times10^{46}{\rm {erg\over s}}}\right)^{{1\over2}} 
\left({T_s \over 1800 {\rm K}}\right)^2 \left({\epsilon\over 0.1}\right)^{-{1\over2}}{\rm cm},
\end{equation}
where $L$ is the QSO luminosity, $\epsilon$ is the dust emissivity,
and $T_s$ is the dust sublimation temperature.
$R_{\rm dust}$ is larger than $R_{\rm E}$ so 
that if dust is responsible for the emission, microlensing fluctuations will
be very small ($\simeq 0.05$ magnitudes for $R_{\rm dust}/R_{\rm E}=30$,
Refsdal and Stabell 1991).

On the other hand, if the infrared emission is due to synchrotron, then
a turnover should be present at low frequencies due to self-absorption.
Such turnovers are typically observed in quasars in the far infrared,
but they are much steeper than the $F_\nu\propto\nu^{5/2}$ behavior
expected from homogeneous self-absorbed synchrotron models (Hughes et al.
1993).  Nevertheless, if such a turnover is due to self-absorption, then 
the requirement that the source be optically thick at the turnover frequency
places a strong upper bound on the size of the emission region of
\begin{equation}
R_{\rm synch}\sim10^{14}\left({F_\nu\over1{\rm mJy}}\right)^{1/2}
\left({B\over 1{\rm G}}\right)^{1/4}\left({\lambda\over100\micron}\right)^{5/4}
\,\,{\rm cm},
\end{equation}
where $F_\nu$ is the flux at the
turnover wavelength $\lambda$, and $B$ is the magnetic
field in the emission region.
Unfortunately the Einstein Cross lies in a region of the sky which was not
surveyed by {\it IRAS}, so the far infrared flux and
turnover wavelength are unknown.  The typical flux in the far-infrared
for a QSO as bright as the Einstein Cross is $\sim$ 10 mJy, so the size of 
the synchrotron emission region is expected to be comparable to the optical size
(where $B$ could be $\sim10^4$~G).
The limit on the size of a synchrotron emitting region is therefore much smaller
than $R_{\rm E}$.  If synchrotron emission is responsible for the IR, then 
microlensing variations in this region of the spectrum will be at least as large 
as in the optical.

A lack of microlensing might allow measurement of the amplification due to the 
gravitational lens galaxy.  
If the IR 
emission is due to dust, then it will not be microlensed, and, as emphasized by 
Kochanek (1991), would provide a direct measurement of the relative magnifications 
due to the lens galaxy rather than due to microlensing (although microlensing by
clusters of stars can still affect a large emission region).   This data is
required for accurate 
modeling of this lens (Witt and Mao 1994).  This has been attempted by using the 
CIII] line, which originates from the extended broad line region and is therefore not
microlensed (Yee \& De Robertis 1992, Racine 1992, Fitte \& Adam 1994, Saust 1994,
Lewis et al. 1998); however, the CIII] line is affected by extinction and continuum 
subtraction. 
To avoid these problems, the flux ratios were measured in the radio by
Falco et al. (1996) with a signal-to-noise ratio (SNR) of only 2-4. One can 
do much better in the IR.  Assuming a power-law mass distribution in the
lensing galaxy ($M\propto r^\beta$), Wambsganss \& Paczy\'nski (1994)
demonstrated that the total magnification of the lens could vary dramatically
as $\beta$ varies.  Their model predicts different flux ratios between
the four images as $\beta$ varies, so measuring the flux ratios of an extended
source might constrain the total amplification.

We present the first observations of the Einstein Cross near 10 microns 
($\sim$3 microns rest frame).  These observations were taken at Keck, which
is the only telescope with a large enough
collecting area to detect such a faint QSO.  In addition, at 10 microns it
is diffraction
limited with about $0\farcs 25$ resolution, making it easy to distinguish
between the four images which have a smallest separation of about \ref{timtab}$\farcs$
In section 2, the observations are discussed.  In section 3, the data
reduction is discussed and the results are presented.  In section 4 the implications 
for emission models and lens models are discussed.  In section 5 the results are
summarized and future directions are pointed out.

\section{Observations}

The Einstein Cross was observed with the Keck Long Wavelength Spectrometer (LWS) 
on three nights in 1999.  The LWS is a 128$\times$
128 Boeing Si:As array  with a $10\farcs5$ field of view at the Cassegrain
focus of Keck I (Jones \& Puetter 1993). 
The LWS was used in imaging mode with chop and nod which were 
necessary due to the faintness of the source and sky variability.  The chop and nod were both
fixed at $10\arcsec$, which is off of the LWS chip.  The field was dithered by $1\arcsec$
in a 5-point pattern.  The 8.9 and 11.7 $\micron$ filters were used, which have 
$\ga 80$\% flux transmission from 8.4 to 9.2 $\micron$ and 11.2 to 12.2 
$\micron$, respectively. Total observing times and on-source integration
times are listed in Table \ref{timtab}.  In addition to observing the QSO, the 
stars HR 5616, 4 Lac, and HR 8551 were observed (near the same airmass) at 
11.7 microns, yielding 13,300 ADU/Jy/sec with a standard deviation of 5\% for different
objects and different nights.  We observed stars HR 8551, $\alpha$ Boo, and HR 2443
at 8.9 microns, yielding 10,300 ADU/Jy/sec with a standard deviation of 6\%.
\vskip 1mm
\hbox{~}
\centerline{\psfig{file=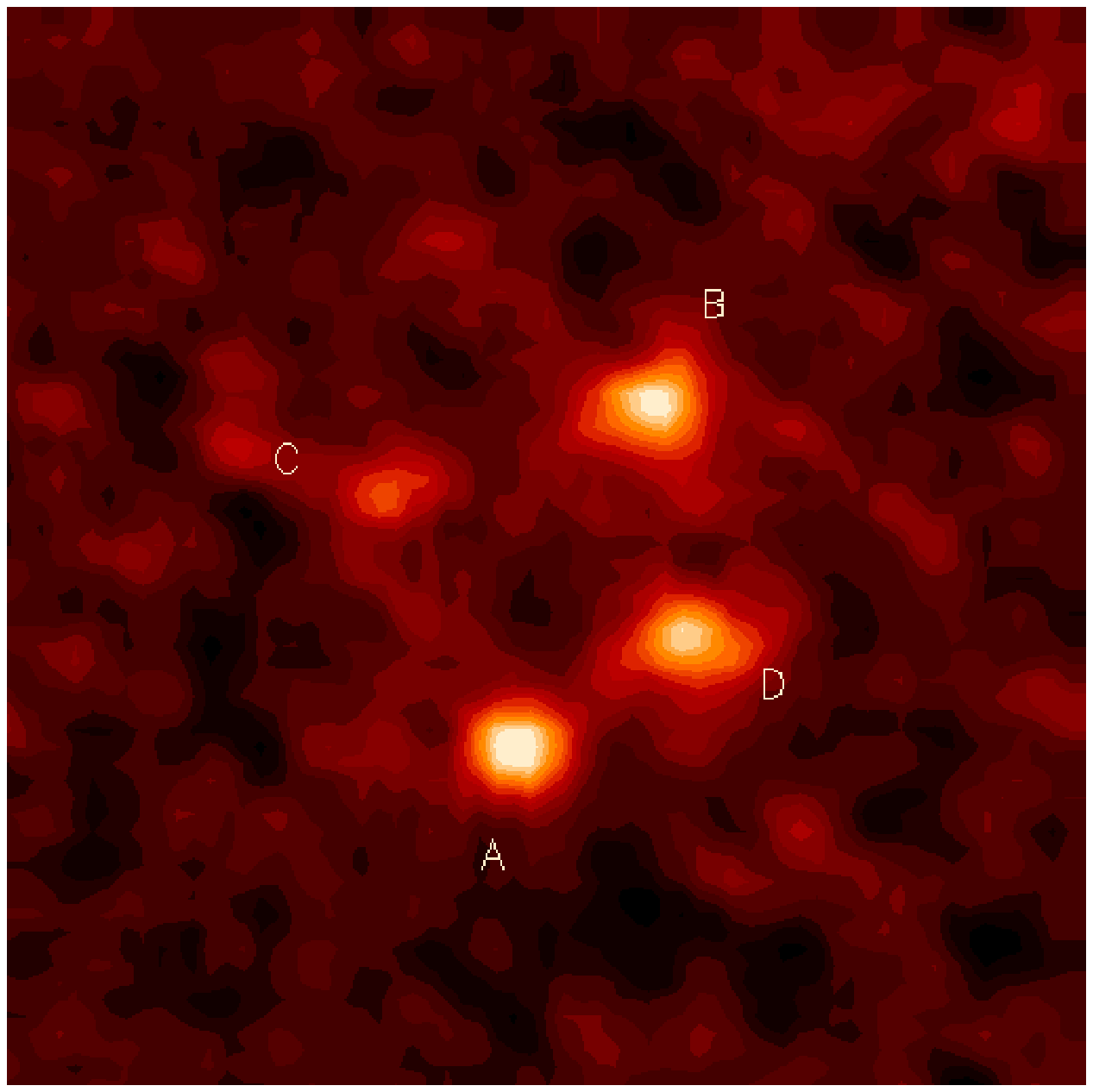,width=3.6in}} 
\noindent{
\scriptsize \addtolength{\baselineskip}{-3pt}
\vskip 0.5mm
\begin{normalsize}
Fig. 1: $5\arcsec\times5\arcsec$ image of the Einstein Cross at 8.9 
and 11.7 microns taken with LWS on July 28, 1999 and September 25 \& 26 
1999 (all data summed together; North is up).
The images are labeled according to Yee (1988).
bottom-right image.
\end{normalsize}
\vskip 1mm
\addtolength{\baselineskip}{3pt}
}
\begin{deluxetable}{cccc}
\tablewidth{6.in}
\tablecaption{Observing times}
\tablehead{ \colhead{Date}
& \colhead{Filter}
& \colhead{Observing time}
& \colhead{On-source time}}
\startdata
7/28/99  & 11.7 $\micron$ & 50  min & 15 min \nl
7/28/99  & 8.9  $\micron$ & 43  min & 12 min \nl
9/25/99  & 11.7 $\micron$ & 157 min & 45 min \nl
9/26/99  & 8.9  $\micron$ & 60  min & 16 min \nl
\enddata \label{timtab}
\end{deluxetable}

\section{Reduction}

\subsection{Basic Reduction}

Successive chopped frames were subtracted and added to successive nod sets 
to get each frame.  Frames were median clipped to get rid of cosmic rays 
and bad pixels.  As the QSO is rather faint, it is not detected in any 
single frame, which typically consists of about five minutes of integration.
To detect the QSO, the data were de-dithered and successive frames were added.
Due to imperfect pointing, the de-dithering can cause some broadening of
the point-spread-function (PSF).  As the QSO is too faint to allow
correction of pointing errors for individual frames, 
the stellar PSF was convolved with gaussians of various widths to see how
this affects the results, as discussed below.  The standard star PSF was
constructed using an identical de-dithering procedure as that applied to
the QSO images, but also was corrected for pointing errors.  
Scaled and broadened stellar PSFs were fitted and subtracted using the known 
relative QSO image positions
(from Falco et al. 1996) while varying absolute position on the chip to 
minimize the residuals of the image minus the model using a linear
least-squares fit.   Using the relative positions measured by HST
(Rix et al. 1992) changed the flux ratios by an amount much
smaller than the error bars.
The resulting fluxes are shown in Table \ref{tab1}.
Figure 1 shows the summed image from both bands and both nights,
smoothed with a 0.25$\arcsec$ boxcar to eliminate noise.
The lens galaxy is undetected; neither is there any evidence of a fifth lensed QSO image.

For comparison to models, the fraction of the total Einstein
Cross flux contained in each image was computed, hereafter referred to as the 
``flux ratio"; i.e. if all four images have the same flux, then the
flux ratio is 1/4 for each image.

\subsection{PSF}

As mentioned above, the pointing errors resulting from dithering can lead 
to a broadening of the PSF for long observations on faint sources, such as 
those presented here.  A stellar PSF (corrected for pointing errors) was 
convolved with a gaussian, varying the FWHM until the best fit
was obtained for each image.  The typical FWHM varies between 3 and 5 pixels
(see Table \ref{tab1}), but the flux ratio of each image on each night in each waveband 
varies by only 5\% if the FWHM is changed from 3 to 5 pixels ($0\farcs25$ to
$0\farcs4$).
This demonstrates that the {\it relative} flux ratios are rather 
insensitive to the gaussian FWHM.  The absolute fluxes, however, are quite 
sensitive to the assumed FWHM.  Fortunately, the argument for microlensing 
depends only on the relative flux ratios.
\begin{deluxetable}{ccccc}
\tablewidth{6.in}
\tablecaption{Infrared fluxes }
\tablehead{ \colhead{Image} &\colhead{\parbox{1.5cm}{Flux 8.9$\micron$ (mJy)}} 
& \colhead{\parbox{1.5cm}{Flux 11.7$\micron$ (mJy)}}  
&\colhead{\parbox{1.5cm}{Flux 8.9$\micron$ (mJy)}} 
& \colhead{\parbox{1.5cm}{Flux 11.7$\micron$ (mJy)}} }
\startdata
Date& 7/28/99     & 7/28/99     & 9/24/99    & 9/24/99     \nl 
FWHM (pixels) & 4.5$\pm$0.6 & 3.1$\pm$0.8 & 4.5$\pm$0.6& 4.9$\pm$0.6 \nl 
\hline \nl
A   & 6.4$\pm$1.2 & 5.7$\pm$1.0 & 6.5$\pm$1.4& 6.4$\pm$1.5 \nl
B   & 9.1$\pm$1.2 & 6.1$\pm$0.9 & 5.7$\pm$1.3& 5.8$\pm$1.3 \nl
C   & 4.2$\pm$1.0 & 3.5$\pm$0.7 & 3.3$\pm$1.2& 2.6$\pm$1.2 \nl
D   & 7.0$\pm$1.1 & 6.2$\pm$0.9 & 6.8$\pm$1.4& 4.2$\pm$1.4 \nl
All &26.8$\pm$2.7 &20.5$\pm$2.1 &22.3$\pm$3.3&18.9$\pm$3.6 \nl
\enddata \label{tab1}
\end{deluxetable}

\subsection{Error bars}

A Monte Carlo technique was used to compute the error bars on the fluxes.
During these observations the LWS chip exhibited pattern noise which prevented
a direct measurement of the error bars from photon counting statistics.  Instead, 
the best-fit model of the four images was added back in at various points on the 
chip to create simulated images.  The simulated images were run through
the entire reduction procedure, measuring the position, gaussian FWHM, and fluxes of
the four images.  This was repeated $\sim$100 times to compute the standard deviation
of the image fluxes, flux ratios, FWHM, and position.  The error bars
on the flux ratios are typically about a factor of 2 greater than what one would 
infer from shot-noise statistics only.  Thus, only the Monte Carlo error bars are 
reported in this paper (Table \ref{tab1}; 1-sigma error bars are always quoted).  The
error bars on the absolute flux are quite large, mostly due to the uncertainty in the
point spread function.  The error bars on the measured FWHM are reported in Table 
\ref{tab1}; this is the FWHM gaussian which was convolved with the stellar PSF to 
approximate the QSO PSF.

\subsection{Consistency of the data}

The flux ratios measured on both nights and in both bands are consistent with 
being constant with time and with wavelength, as shown in Figure 2.  Any disagreement 
between the measurements are consistent with random errors causing the differences;
there is no need to invoke variability in the flux ratios with time or between
the two bands, nor any need to find systematic errors to explain the differences.

\vskip 2mm
\hbox{~}
\centerline{\psfig{file=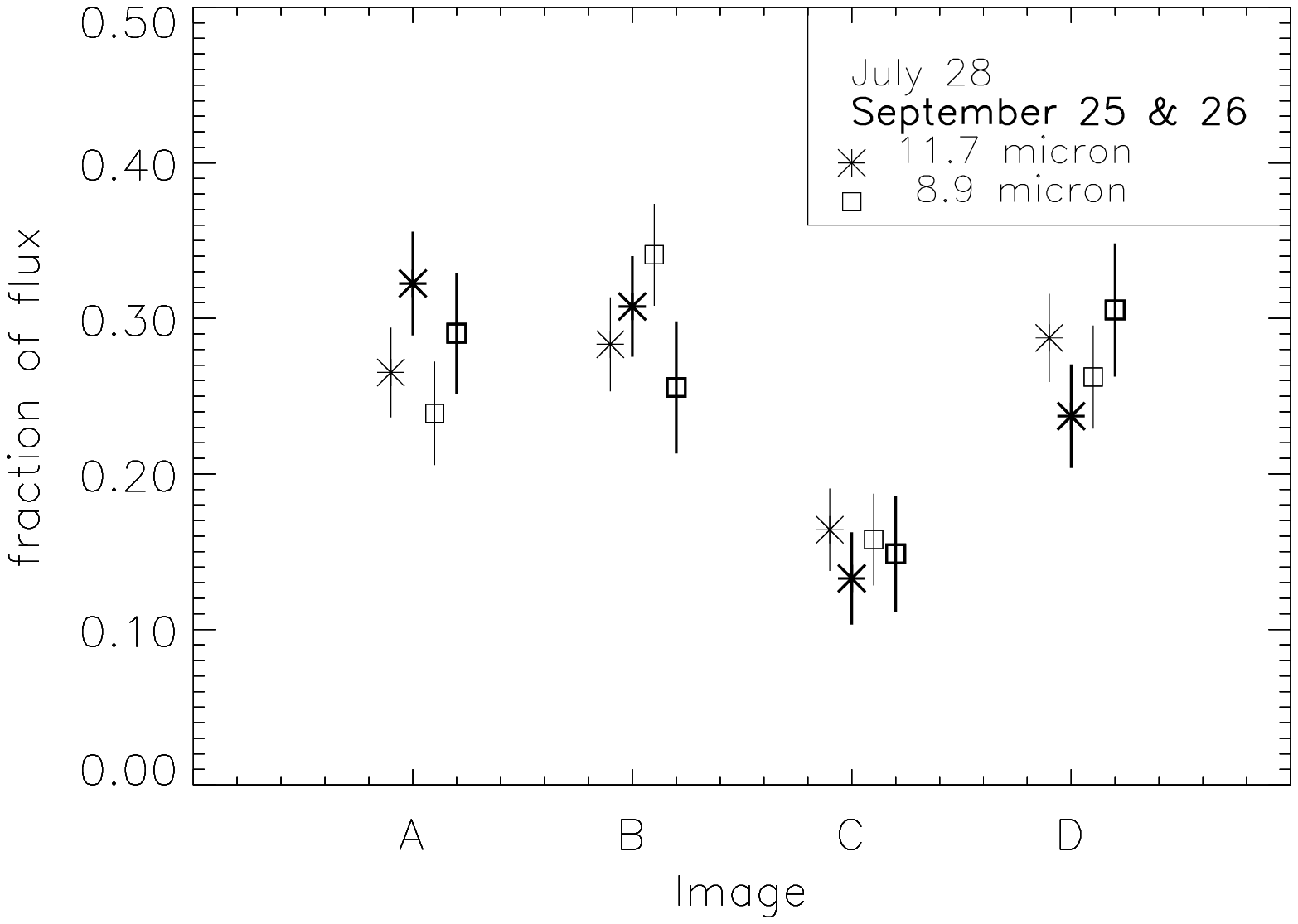,width=3.6in}} 
\noindent{
\scriptsize \addtolength{\baselineskip}{-3pt}
\vskip 1mm
\begin{normalsize}
Fig. 2:  Fraction of the total flux of the Einstein Cross
contained in each image.  The data from September are plotted in bold;
the 11.7 $\micron$ data with asterisks; and the 8.9 
$\micron$ data with squares.  The error bars reflect the statistical uncertainties
\end{normalsize}
\vskip 3mm
\addtolength{\baselineskip}{3pt}
}

\section{Discussion}

\subsection{Microlensing Constraints}

Table \ref{tab2} compares the mid-infrared flux ratios (averaged from both
nights and both bands) to the flux ratios in other wave bands.
The V band magnitudes are from the OGLE monitoring data on the dates closest to 
the Keck observations (available on their website; Wo\'zniak et al. 2000);  
this data was corrected for extinction as described in the Appendix.
The VLA radio data are from Falco et al. (1996) which were
taken in 1995.  The CIII] ratios are from Racine (1992), data taken with CFHT in 
1991.  One can immediately draw the conclusion that the infrared is extended
(and thus probably dust emission) from the following lines of evidence:

1)  The flux ratios differ between the IR \& V band by more than 6$\sigma$ 
for each image.  Simultaneous with the Keck observations were strong optical 
microlensing events in images A and C at the time of the infrared observations
(OGLE website; Wo\'zniak et al. 2000).  Since the optical is known to be
microlensed, this 
indicates that the IR is extended on a scale comparable to or larger than an 
Einstein radius, causing dramatic differences in the flux ratios.

2) The image ratios in the IR are close to the ratios predicted by some lens 
models, lending weight to the idea that the IR is more extended and thus not 
affected by microlensing.  Table \ref{tab2} shows the predictions of
the $\lambda=1$ model from Schmidt, Webster, \& Lewis (1998, hereafter SWL), 
which were derived {\it without} using the flux ratios as a model constraint.
For all published models $\chi^2 =\sum_{i=\{A,B,C,D\}} 
(f_i(obs)-f_i(mod))^2/\sigma_i^2$ was computed, where $f_i$ is fractional flux of each image and 
$\sigma_i$ is the measured error bar.  The three best models are $\lambda = 1$ of SWL 
with $\chi^2=7$,  model 4 of Kochanek (1991) with $\chi^2=7$, and model 1 of
Witt, Mao, \& Schechter (1995) with $\chi^2=8$.  These $\chi^2$ indicate that there 
may be deviations of the infrared magnifications from the predicted magnifications at the 
level of 10\%.   This may be due to inaccuracies in the modeling, an underestimation
of the error bars, or microlensing fluctuations at the 10\% level, which is to be
expected for a source 15 times the size of an Einstein radius (Refsdal \& Stabell
1991, assuming $\sigma = 0.5$).  
All other published gravitational lens models for Einstein Cross have
larger $\chi^2$ (Kent \& Falco 1988; Kochanek 1991; Rix, Schneider, \& Bahcall 1992; 
Wambsganss \& Paczy\'nski 1994; Witt et al. 1995; Chae, Turnshek, \& Khersonsky 1998).

3)  The image ratios at 8.9 and 11.7 microns are consistent within the
error bars, indicating that neither wavelength is affected by either extinction or
differing microlensing amplification due to different sizes as a function of
wavelength.

4)  The radio emission is thought to be extended on a scale larger than an 
Einstein radius (possibly in a jet) and thus is unaffected by microlensing.  
Since the IR image ratios are consistent with the radio image ratios within 
the error bars, this implies that the IR may also be unaffected by microlensing.  
Note that the infrared SNR is much better than in the radio.

5)  In the case of images C and D, the IR image ratios are not
consistent with the CIII] image ratios, but agree better than with the
V band flux ratios.  The deviations may be due to microlensing fluctuations 
of CIII] if it is only a few times larger than an Einstein radius, or due
to the systematic errors which affect the CIII] measurements.

6)  Image C dropped in optical flux by a factor of 1.4 between August 4 and
September 26 (the two OGLE observations closest in time to the two Keck
observations).
Our observations are consistent with no variation in image C in either
band between the two dates, although the possibility of a variation is not
strongly ruled out.

7) Consider a model in which a point source and an extended source contribute
to the infrared flux (which would be the case if both a dusty torus and
synchrotron source contributed to the infrared flux).  Then, the synchrotron
source would be magnified by the same amount as the optical (due to 
microlensing), while the extended source would be magnified by the
macrolensing magnification (assuming the SWL model holds).
With these assumptions one can compute what the fraction of point source flux 
is relative to the extended source flux in order to give the infrared flux
ratios.  Carrying out this exercise, one finds that the point source contribution
must be {\em negative} to explain the difference between images A and B or
A and C.  Simply put, the infrared flux ratio of B is larger than the model,
while the point source (i.e. optical) is smaller than the model (see Table 
\ref{tab2}), so adding a point source to an extended source can only decrease 
the flux with respect to the model.  Since the point source flux 
must be non-negative, it is most likely zero. Then, the disagreement between the 
infrared flux ratios and the model flux ratios is probably due to inaccuracies
in the lens model, Poisson fluctuations due to a random number of lenses
passing in front of a given image  (since the source is not truly infinite),
and/or errors in the measured fluxes.

Given these lines of evidence, one concludes that the mid-infrared emission
region is larger than the Einstein radius of a typical star in the lens
galaxy, about $10^{17}$ cm.  The infrared may be extended by $0\farcs24$
which corresponds to about 0.8 kpc in the source plane; however, this
extension is possibly due to pointing errors as discussed above.
Consequently, $0.03 {\rm pc} \la R_{IR} \la 800 {\rm pc}$.
Comparing the optical and infrared data with
simulated microlensing lightcurves with various source sizes, one can compute the 
probability distribution of the infrared source size, which will be done in future 
work (Wyithe \& Agol, in preparation).

\vskip 2mm
\hbox{~}
\centerline{\psfig{file=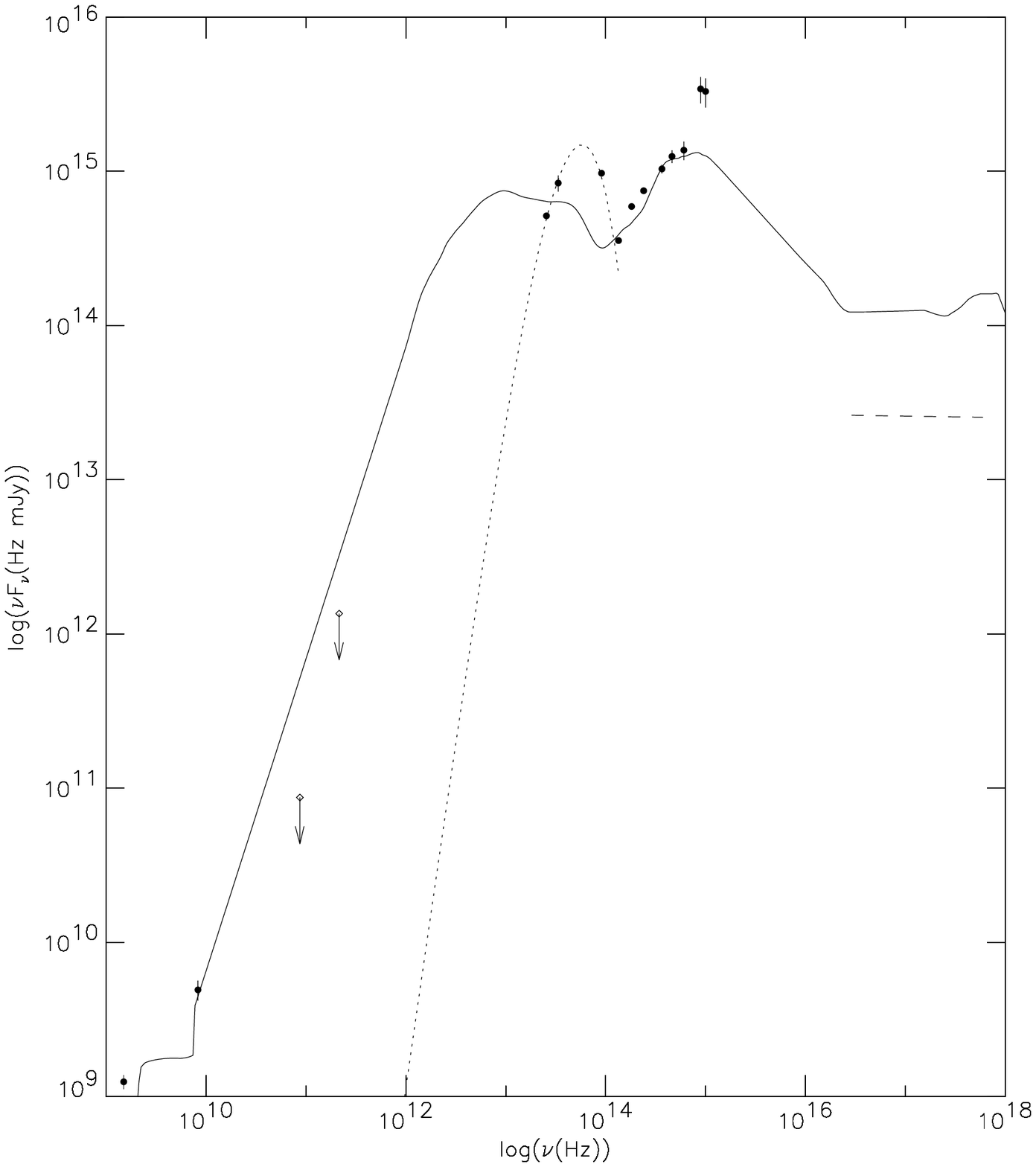,width=3.6in}} 
\noindent{
\scriptsize \addtolength{\baselineskip}{-3pt}
\vskip 1mm
\begin{normalsize}
 Fig. 3: Spectral energy distribution of
the Einstein Cross (non-simultaneous; frequency in observed frame).
The near-infrared and optical points are taken from Nadeau et al. (1991);
the ultraviolet from Blanton, Turner, \& Wambsganss  (1998); the X-ray data (dashed line) is
from Wambsganss et al. (1999); the radio points are from Falco et al.
(1996); the mm points (diamonds) are 3 $\sigma$ upper limits from Richard Barvainis 
(priv. comm.); and the mid-infrared data are from this work. The solid line is the
radio-quiet QSO composite from Elvis et al. (1994) scaled to match the
optical flux. The dotted line is the dust shell model described in the
text based on Barvainis (1987).
\end{normalsize}
\vskip 3mm
\addtolength{\baselineskip}{3pt}
}

\begin{deluxetable}{ccccccc}
\tablewidth{6.in}
\tablecaption{Flux fractions of images}
\tablehead{\colhead{Wavelength} & \colhead{Date} & \colhead{Image A} 
& \colhead{Image B} & \colhead{Image C} & \colhead{Image D} }
\startdata
8.9 \& 11.7$\micron$ &  7-28 \&  9-24  &  0.27$\pm$0.02 & 0.30$\pm$0.02 &  0.16$\pm$0.02 &0.27$\pm$0.02 \nl
V-band\tablenotemark{a} &   8-4-99& 0.39 & 0.11 & 0.41 &0.10  \nl
V-band\tablenotemark{a} &  9-26-99& 0.46 & 0.12 &0.32 &0.10  \nl
8 GHz & 1995          & 0.29$\pm$0.06 &0.32$\pm$0.16 &0.16$\pm$0.06 &0.22$\pm$0.06 \nl
CIII] & 1991 &  0.26$\pm$0.01&0.29$\pm$0.01&0.26$\pm$0.01& 0.19$\pm$0.01\nl
Model (SWL) & & 0.25 & 0.27 & 0.15 & 0.32 \nl
\enddata \label{tab2}
\tablenotetext{a}{Corrected for extinction}
\end{deluxetable}

\subsection{Spectral Energy Distribution}

To study in more detail the properties of the Einstein Cross, an SED was 
constructed of all available data from the radio to X-ray from
VLA, IRAM, CFHT, Keck, HST, and ROSAT.  The lens galaxy has been
removed (or does not contribute) for all the observations shown.
The near-IR, optical, UV, and X-ray were corrected for extinction in the
lens galaxy and the Milky Way as described in the Appendix.
This brings the optical/UV slope into rough agreement with the average QSO
slope.  Figure 3 shows the SED in the observed frame with the extinction
corrections, but without a correction for magnification.  Figure 3 also shows
the Elvis et al. (1994) radio-quiet QSO composite redshifted and scaled to the 
R band flux.  Assuming a magnification of 16 (from the SWL model - this value 
is used throughout the rest of this paper), this implies a QSO luminosity
of $2\times 10^{46}$ erg/s for the scaled composite.  However, there appear to
be deviations of the Einstein Cross from the composite QSO in the mm, infrared, UV, and 
X-ray.  The mm upper limits are slightly lower than the average SED.  The infrared 
has a narrower peak at higher frequency than the average QSO.  A similar peak
exists in 3C 273; however, 3C 273 peaks at 3 $\micron$ and is broader than
the Einstein Cross (Barvainis 1987).  The ultraviolet points are high, which may mean that 
this is a hot QSO or that the QSO has varied (or been magnified by microlensing)
as the UV points were taken nearly a 
decade after the optical/near-IR points.  The X-rays fall a factor of 3 below 
the average QSO, which is within the large scatter of optical to X-ray flux 
ratios of QSOs.  The radio to optical flux ratio is identical to the QSO 
composite.  The infrared discrepancy is of greatest relevance to this paper, 
which is discussed in turn.

\subsection{Origin of Infrared Bump}

The Keck mid-IR fluxes (summed over all four images) give a spectral index 
of $\alpha = 0.5 \pm 0.2$ ($F_\nu \propto \nu^\alpha$), which is quite 
different from the average for PG QSOs, $\alpha=-1.4\pm0.5$ (Neugebauer 
et al. 1987).  This unusual behavior may be due to calibration problems.
However, we think our error bar estimates are accurate, and we are unable to identify
any calibration errors which affect the Keck observations at the 50\% level
which would be required to bring the slope into agreement with that of an
average QSO.  The data were compared with ISO CAM observations (TDT 54500601;
P.I. Sangeeta Malhotra) of the Einstein Cross from the ISO archive which 
show a similar infrared slope of $\alpha=0.45$.  The ISO CAM fluxes at 9.1 
and 11.7 microns agree with the LWS measured fluxes within the 1$\sigma$
error bars.  The near-IR spectral index is dramatically different:  
Nadeau et al. (1991) report $\alpha = -3.7$ between the K-band and L-band
(3.3 micron band).  This implies that there is a peak in $\nu F_\nu$ between 
3.3 microns and 8.9 microns (1.2 microns and 3.3 microns in the rest frame); 
i.e. the peak infrared flux occurs somewhere near 2 microns.  Typically, QSOs peak
much further in the infrared, near 10 microns (see Figure 3).

This unusual infrared bump might be explained as a quasi-blackbody component.   
There are several candidate physical components which might
contribute near these wavelengths:  synchrotron, hot dust, a starburst, or the host
galaxy.  Synchrotron is ruled out by microlensing, so the other
possibilities are considered next.

\subsubsection{Hot Dust}

A blackbody component was fit at the dust sublimation temperature
($\sim2000$~K), 
which agrees with the data remarkably well if the effective radius (area = $\pi r_d^2$) 
of the source is $r_d = 2\times 10^{18}$~cm.
Treating the dust emission as a blackbody assumes
the emissivity is unity, appropriate for large grains; small grains would reduce
the emissivity, requiring a larger area, but lowering the temperature to 1600 K
(for $\epsilon = 2\pi r/\lambda$, where $r \ll 1\micron$ is the dust grain radius).
The inferred source size agrees with the dust sublimation estimate {\it and} with 
the microlensing size constraint.  However, this implies that the dust 
should be located at a constant radius so that when it is exposed to the QSO continuum 
it equilibrates at a constant temperature.  This means that the dust is in a shell
geometry;  if there is dust at a range of radii, then it will radiate
at lower temperatures than the sublimation temperature which will make 
the infrared SED wider and flatter than a blackbody, typically rising in
flux towards longer wavelengths (Pier \& Krolik 1992).  

To quantify the constraint on geometry, the near-infrared bump was fit with 
the spectral model of 
Barvainis (1987).   Barvainis' model assumes that the dust is in a shell
subtending $\Omega$ steradians of the central source with an inner radius
($r_1$) set by the luminosity of the source ($L$) and the dust sublimation
temperature ($T_{\rm max}$).  The shell is assumed to be optically thin in
the near-IR ($\lambda \ga 1\micron$), with a UV optical depth of 3.
Barvainis' uniform dust model was adopted with $a=0.05 \micron$, emissivity
proportional to $\nu^{1.7}$, and we additionally 
assume that the dust grains have a constant number density, that 
$L=2\times 10^{46}$ erg/s, and that $T_{\rm max}=1800$K, fixing the inner radius 
at 1.1 pc, consistent with the microlensing size constraint.  The two free 
parameters left are $r_2/r_1$ (the ratio of the outer to inner radius of 
the shell) and $\Omega$.  An adequate fit was found for $r_2/r_1 = 3$
and $\Omega = 7\pi/4$.   In this model, the dust outside $r_2$ is
shielded, and so does not contribute to the reprocessed light.
The grain number density at the inner edge is $5\times 10^9$ cm$^{-3}$, so
the implied dust mass is about $4 {\rm M_\odot}$ for a dust density of 1 g/cm$^3$.
The Toomre Q-parameter is quite small, $\sim 10^{-9}$ (for a mass of
$2\times 10^9 M_\odot$),  which indicates that this region should be 
gravitationally unstable.  A clumpy disk model should have a similar spectrum
since the smaller filling factor can be compensated by an increase in dust 
number density.  The conclusion is that the narrow infrared peak requires a
narrow range of radii (factor of 3) so that reprocessed luminosity
is absorbed by a region with a narrow range of dust temperature.
This narrow range is not unprecedented as Barvainis (1992) found a narrow
range of radii for the dust emission in Fairall 9 from reverberation
$r_1 \la 0.3$pc and $r_2 \sim 1.3$pc.

This, of course, is a rather simplistic model.  X-rays may penetrate
deeper, heating dust at larger radii.  Transient heating by X-rays causes
temperature flickering which broadens the spectrum;  this is only important 
for dust grains with radii $a \la 0.01\micron$ (Voit 1991).  
Infrared radiation transfer has been ignored since the shell is optically 
thin by construction; reality may not be so simple.  A warped disk with
a warp angle which scales logarithmically radius would produce a much
wider infrared bump than observed.

\subsubsection{Host Galaxy or Starburst}

Normal galaxies can be well described by a 3500 K blackbody in the near infrared 
spectral region (Schmitt et al. 1997), which peaks at a wavelength of about 
$1 \micron$ (rest), while the Einstein Cross spectrum peaks closer to $2 \micron$ 
(rest).  Thus, to fit the infrared bump 
requires strong extinction as in a starburst galaxy:  an extinction of $A_V\sim 7$ 
is needed to fit the infrared bump; most of the light is absorbed by dust.   
The absorbed luminosity is $L_{\rm abs}=6\times 10^{46}$ erg/s, which must be 
re-radiated in the far-infrared (FIR).  The radio-FIR relation (Condon 1992) limits 
the FIR luminosity 
to $6\times 10^{45}$ erg/s, a factor of 10 times smaller than the absorbed luminosity.
A test of the host galaxy hypothesis would be to look for an extended source in the
IR images.  Meurer et al. (1997) have found an empirical limit on the
bolometric surface brightness of starburst galaxies of $2\times 10^{11}{\rm L}_\odot/
{\rm kpc}^2$, which would indicate that this starburst has a size of $\ga$ 3.5 kpc,
or $\sim1\arcsec$ (including the factor of $\sim 2$ stretch due to lensing).  This is 
larger than the infrared upper lmit of $0\farcs25$, ruling out a starburst as the 
origin of the infrared bump.

One can roughly estimate the host galaxy luminosity from the QSO luminosity as 
follows.  Rauch \& Blandford (1991) have argued that sub-Eddington accretion 
disk models are too large to be consistent with the optical size constraint 
from microlensing.  We confirmed this by fitting the optical/UV SED with
various thin accretion disk models from Hubeny et al. (2000); all disks are
a factor of a few too big to be consistent with the microlensing constraint.  
A smaller emission region might be achieved by decreasing 
the mass of the black hole; however, once the accretion rate reaches Eddington, 
the luminosity will be roughly fixed at the Eddington rate.  Thus, if one assumes 
that the observed (isotropic) luminosity equals the Eddington rate, then the black 
hole mass is about $2\times 10^9 {\rm M_\odot}$.  Using the Magorrian et al. 
(1998) relation between black hole mass and bulge luminosity, the bulge luminosity 
is estimated to be about $6\times 10^9 {\rm L_\odot}$, with an uncertainty of
about a factor of 10.  This is more than 150 times smaller than the near-IR
luminosity, yet another indication that the infrared bump may not be due to a 
quiescent host galaxy.

\section{Conclusions}

The observations presented here indicate the mid-infrared source size in the
the Einstein Cross must be greater than $10^{17}$cm, ruling out synchrotron
emission.   If true for other radio-quiet QSOs, this implies that fits
to the Big Blue Bump which include a synchrotron power law may be flawed
(Sun \& Malkan 1989, Laor 1990).
Starburst and host galaxy contributions to the mid-infrared emission are ruled
out based on the low radio flux, small size, and large infrared luminosity,
leaving hot dust as the only viable model for
the infrared emission in the Einstein Cross.  To avoid sublimation,
the dust must be at a distance of about 1 pc from the QSO;
to explain the narrowness of the infrared bump it must be located within
$\sim$ 3pc (according to a simple reprocessing model); and to explain its strength, 
the dust must intercept about half of the QSO continuum.  Since the mid-IR is extended,
the flux ratios should be nearly equal to those of the galaxy gravitational lens
model, so future models should be constructed with the infrared flux ratios
as model constraints.  Improved microlensing statistics, magnifications, 
and time delays will result from better lens models.
Accurate monitoring of the Einstein Cross at mid-infrared wavelengths with, 
say, NGST, may reveal flux variations at the few \% level, which would allow a
better estimate of the size of the dust emission region.  The Einstein Cross
mid-infrared spectrum is quite different from any other QSO;  it should be 
observed with SIRTF to better constrain the dust emission model.

\acknowledgments

Data presented herein were obtained at the W.M. Keck Observatory, which is 
operated
as a scientific partnership among the California Institute of Technology, the 
University of California and the National Aeronautics and Space Administration.
The Observatory was made possible by the generous financial support of the 
W. M. Keck Foundation.  This work was supported by NSF grants AST-9529230
and AST-9970827 and California Space Institute grant CS-16-93.
We would like to thank Greg Wirth and Randy Campbell
for their expert assistance during our observing runs.  We also thank Robert
Antonucci for inspiring this project.  We thank Robert Antonucci, Tim Heckman, 
Julian Krolik, Sangeeta Malhotra, Nancy Levenson, Henrique Schmitt, Mark Voit,
and Rachel Webster for useful discussions.  We thank Richard Barvainis for
sharing his millimeter upper limits before publication.  We thank Przemyslaw 
Wo\'zniak and the OGLE team for making their data available on the web before 
publication.   We have also used archival data from ISO, an ESA project 
with instruments funded by ESA Member States (especially the PI countries: 
France, Germany, the Netherlands and the United Kingdom) and with the 
participation of ISAS and NASA.

\appendix

\section{Appendix: Correction for Extinction}

To correct for extinction in the lens galaxy, the correlation between 
$g-i$ color (Gunn bands) and lens galaxy surface brightness found by Racine (1991)
was used.
Extrapolation to zero surface brightness then gives the color corrected for extinction 
in the lens galaxy, if one assumes that the dust column is proportional to surface
brightness (this assumption may be invalid if the dust is patchy).  To solve for the 
absolute extinction, we assume the extinction curve is described by $R=3.1\pm 0.9$, 
the average Milky Way extinction curve.  $A_V$ for image
A is 0.88$\pm$0.21, B is 0.84$\pm$0.20, C is 1.30$\pm$0.31, and D is  1.15$\pm$0.27.
Using the Milky Way extinction $E(B-V)=0.071$ measured by Schlegel, Finkbeiner, 
\& Davis (1998), the Nadeau et al.  spectrum was corrected for each image for galactic 
extinction using the $R=3.1$ extinction curve (Fitzpatrick 1998).  It is reassuring 
that the corrected spectrum is similar to that of an average QSO
(see Figure 3).  Yee (1988) and Falco et al. (1999) correct for extinction
assuming the Einstein Cross has a spectral index equal to the average value 
for QSOs; since individual QSOs do not behave as the average, we prefer the 
Racine technique.

To correct the X-ray data for absorption, we adopt the dust-to-gas ratio
given by Spitzer (1978):  $N_H=5.9\times 10^{21} E_{B-V} {\rm mag^{-1} cm^{-2}}$.
An average extinction of 1 for the lens galaxy and 0.2 for the Milky Way implies 
$N_H = 2.5\times 10^{21}{\rm cm^{-2}}$.  The X-ray power law was assumed to be equal
to that of an average QSO and the count rate of Wambsganss et al. (1999) was used
to compute the unabsorbed X-ray flux in the ROSAT band.

\end{document}